\documentclass[graybox]{svmult}

\usepackage{mathptmx}       
\usepackage{helvet}         
\usepackage{courier}        
\usepackage{type1cm}        
\usepackage{makeidx}         
\usepackage{graphicx}        
\usepackage{multicol}        
\usepackage[bottom]{footmisc}

\usepackage{amssymb}
\usepackage{amsmath}
\usepackage{dsfont}
\usepackage{hyperref}
\usepackage{pstricks}
\usepackage{auto-pst-pdf}
\usepackage{comment}
\usepackage{subeqnarray}

\makeindex             

\begin{document}

\title*{Boundary States of the Potts Model on Random Planar Maps}
\author{Max Atkin, Benjamin Niedner and John Wheater}
\institute{Max Atkin \at  Universit\'e Catholique de Louvain, Chemin du cyclotron 2, B-1348 Louvain-La-Neuve, Belgium,\\ Fakult\"at f\"ur Physik, Universit\"at Bielefeld, Postfach 100131, D-33501 Bielefeld, Germany, \email{matkin@physik.uni-bielefeld.de}
\and Benjamin Niedner and John F. Wheater \at Rudolf Peierls Centre for Theoretical Physics, University of Oxford, 1 Keble Road, Oxford OX1 3NP, UK, \email{benjamin.niedner@physics.ox.ac.uk}, \email{j.wheater@physics.ox.ac.uk}}
%
%
\maketitle

\abstract*{We revisit the 3-states Potts model on random planar triangulations as a Hermitian matrix model. As a novelty, we obtain an algebraic curve which encodes the partition function on the disc with both fixed and mixed spin boundary conditions. We investigate the critical behaviour of this model and find scaling exponents consistent with previous literature. We argue that the conformal field theory that describes the double scaling limit is Liouville quantum gravity coupled to the $(A_4,D_4)$ minimal model with extended $\mathcal{W}_3$-symmetry.} 


\abstract{We revisit the 3-states Potts model on random planar triangulations as a Hermitian matrix model. As a novelty, we obtain an algebraic curve which encodes the partition function on the disc with both fixed and mixed spin boundary conditions. 
We investigate the critical behaviour of this model and find scaling exponents consistent with previous literature. We argue that the conformal field theory that describes the double scaling limit is Liouville quantum gravity coupled to the $(A_4,D_4)$ minimal model with extended $\mathcal{W}_3$-symmetry. 
}

\section{Introduction}
\label{introduction}

\noindent Matrix models provide a powerful tool for defining quantum gravity partition functions in two dimensions. The conformal field theories describing their critical points generically correspond to Euclidean quantum gravity interacting with conformal matter \cite{DiFrancesco93}
. The spectrum of the theory is then determined by the consistent boundary states and as a result, the computation of such boundary states using matrix models has been of continued interest \cite{Atkin10,Oh11,Chan11,Atkin12}. In this contribution, we shall consider this problem for the 3-states Potts model coupled to gravity, whose partition function can be written as the $U(N)\times\mathbb{Z}_3$-invariant matrix integral

\begin{equation}
	\label{partitionfct}
	Z(N,c,g)=\int\prod_{i=1}^3\D X_i\ \E^{-N\mathrm{tr}[\sum_{i=1}^3V(X_i)-\sum_{\langle ij\rangle}X_iX_j]}\;,
\end{equation}

\noindent where we have set $V(x)=cx^2/2+gx^3/3$ and the $X_i$ are $N\times N$ Hermitian matrices
.
This matrix model was first partially solved in \cite{Daul94}, followed by the more detailed investigations \cite{Bonnet99,Eynard99,ZJ99}; related results have been obtained from combinatorial approaches \cite{Bernardi11,Borot12}. 

To find the relevant observables in this matrix model, we note that in flat space, the critical point of the 3-states Potts model is described by the $(A_4,D_4)$ minimal model, the simplest model in which the presence of an additional higher-spin current extends the Virasoro algebra to the $\mathcal{W}_3$-algebra. Its conformally invariant boundary states were classified in \cite{Cardy89,Affleck98} and are listed in Table \ref{tab:states} together with their discrete counterparts. Accordingly, we introduce the generating function for triangulations of the disc with fixed color $i$ on the boundary, 

\begin{equation}
	\label{discfct}
	w_i(x)=\frac{1}{N}\sum_{k=0}^{\infty}\langle\mathrm{tr} \ X_i^k\rangle x^{-k-1}=\frac{1}{N}\left\langle\mathrm{tr}\frac{1}{x-X_i}\right\rangle\;.
\end{equation}

\noindent We see from  Table \ref{tab:states} that the continuum limit of these observables should correspond to the 3 states $\mathds{1}$, $\psi$ and $\psi^{\dagger}$. The mixed-color boundary conditions $\varepsilon$, $\sigma$ and $\sigma^{\dagger}$ may be imposed by considering
\begin{equation}
	\label{w+}
	w_+(x)=\frac{1}{N}\left\langle\mathrm{tr}\frac{1}{x-(X_i+X_j)}\right\rangle\;,\qquad i\neq j\;.
\end{equation}

\begin{table}[b]
	\label{tab:states}
	\caption{Boundary states in the $(A_4,D_4)$ minimal model, their decomposition in $(A_4,A_5)$ Virasoro modules, their conformal weights $h$, their $\mathbb{Z}_3$ charge $q$, and the corresponding microscopic boundary conditions}
	\centering
	\begin{tabular}{llllll}
		\hline\noalign{\smallskip}
		&$(A_4,D_4)$&$(A_4,A_5)$&$h$&$q$&\\
		\noalign{\smallskip}\svhline\noalign{\smallskip}
		$\mathds{1}$&$(1,1)$&$(1,1)\oplus(1,5)$&$0,3$&$0$&$X_1$\\
		$F$&$(1,2)$&$(1,2)\oplus(1,4)$&$1/8,13/8$&--&$X_1+X_2+X_3$\\
		$\psi$&$(1,3)$&$(1,3)$&$2/3$&$+1$&$X_2$\\
		$\psi^{\dagger}$&$(1,4)$&$(1,3)$&$2/3$&$-1$&$X_3$\\
		$\varepsilon$&$(2,1)$&$(2,1)\oplus(3,1)$&$2/5,7/5$&$0$&$X_2+X_3$\\
		$N$&$(2,2)$&$(2,2)\oplus(2,4)$&$1/40,21/40$&--&\\
		$\sigma$&$(2,3)$&$(2,3)$&$1/15$&$+1$&$X_1+X_3$\\
		$\sigma^{\dagger}$&$(2,4)$&$(2,3)$&$1/15$&$-1$&$X_1+X_2$\\
		\noalign{\smallskip}\hline\noalign{\smallskip}
	\end{tabular}
\end{table}

\noindent The remaining boundary conditions $F$ and $N$ correspond to operators not contained in the bulk spectrum of $(A_4,D_4)$; these will not be discussed herein. The purpose of this note is then to compute the functions \eqref{discfct} and \eqref{w+} at large $N$\footnote{Note that whilst on the disc, the individual boundary conditions that form an orbit under the $\mathbb{Z}_3$ action are indistinguishable, this will generally not be the case for higher topologies.}. In the next section, we shall demonstrate that this can be achieved by requiring the saddle point equations to have a consistent analytic continuation in this limit. 

\section{Discrete Solution}
\label{discrete}

\noindent In this section, we determine the disc partition functions with boundary conditions corresponding to the orbits $\{{\mathds 1},\psi,\psi^{\dagger}\}$ and $\{\varepsilon,\sigma,\sigma^{\dagger}\}$ under the action of $\mathbb{Z}_3$ in Table \ref{tab:states} for generic values of the couplings $c$ and $g$. Consider the change of variables

\begin{equation}
	\label{newvars}
	X_1=\frac{1}{2}(X_++X_-)-\frac{c+1}{2g}\;,\qquad X_2=\frac{1}{2}(X_+-X_-)-\frac{c+1}{2g}\;,
\end{equation}

\noindent so that the resolvent of $X_+$ coincides with the expression \eqref{w+} for the disc partition function with boundary condition $\sigma^{\dagger}$ up to a shift in $X_+$ which leaves $Z$ invariant. We may pick $w_+(x)$ and $w_3(x)$ as representatives of each $\mathbb{Z}_3$ orbit. In these variables, the integrand in \eqref{partitionfct} is Gaussian in $X_-$, which can hence be integrated out, leaving us with an integral over just two matrices. Upon gauge-fixing the $U(N)$-symmetry, we can then carry out the integral over the unitary group using the well-known result \cite{Itzykson80}. Denoting the respective eigenvalues of $X_+$ and $X_3$ by $x_+^i$ and $x_3^i$, $i=1\dots N$, the resulting saddle point equations read

\begin{subeqnarray}
	\label{saddle}
		\frac{\partial U_+(x_+^i)}{\partial x_+^i}&=&\frac{1}{N}\left(\frac{\partial}{\partial x_+^i}\ln\det_{k,l}\ \E^{Nx_+^kx_3^l}+\sum_{j<i}\frac{1}{x_+^i-x_+^j}-\sum_j\frac{1}{x_+^i+x_+^j}\right)\;,
		\label{saddle1}\\
		\frac{\partial U_3(x_3^i)}{\partial x_3^i}&=&\frac{1}{N}\left(\frac{\partial}{\partial x_3^i}\ln\det_{k,l}\ \E^{Nx_+^kx_3^l}+\sum_{j<i}\frac{1}{x_3^i-x_3^j}\right)\;. 
		\label{saddle2}
\end{subeqnarray}

\noindent Here, $U_+$ and $U_3$ are polynomials of degree $3$ with coefficients determined by $V(x)$ and \eqref{newvars}. Since the left-hand side of \eqref{saddle} is holomorphic, the above equations can be analytically continued to the complex plane for any $N$. On the other hand, for $N\to\infty$, individual expressions on the right-hand side will develop branch cuts located at the support of the densities of eigenvalues. To determine their analytic continuations, we follow \cite{ZJ98} in introducing the functions

\begin{equation}
	\label{fcts}
	x_3(x)=\lim_{N\to\infty}\left.\frac{1}{N}\frac{\partial}{\partial x_+^i}\ln\det_{k,l}\E^{Nx_+^kx_3^l}\right|_{x_+^i=x}\;,
	\quad 
	x_+(x)=\lim_{N\to\infty}\left.\frac{1}{N}\frac{\partial}{\partial x_3^i}\ln\det_{k,l}\E^{Nx_+^kx_3^l}\right|_{x_3^i=x}\;.
\end{equation}

\noindent Let us denote the spectral density of $X_+$ by $\rho_+$, whose support we assume connected
. It was shown in \cite{ZJ98} that for $N\to\infty$,

\begin{enumerate}
	\item $\exists\gamma\in\mathbb{R}$ such that the function $x_3(x)$ is analytic on $\mathbb{C}\setminus(-\infty,\gamma]\cup\text{supp}\ \rho_+$,	
	\label{prop1}
	\item its discontinuity across $\text{supp}\ \rho_+$ coincides with that of $w_+(x)$ and
	\label{prop2}
	\item it is the functional inverse of $x_+(x)$.	
	\label{prop3}
\end{enumerate}

\noindent By the symmetry of the definitions \eqref{fcts}, the analogous statements apply to $x_+(x)$. Furthermore, let $x_3^*(x)$ (resp. $x_+^*(x)$) be the function obtained by analytic continuation through $\text{supp}\ \rho_+$ onto the next sheet. Using properties \ref{prop1} and \ref{prop2} above, the large $N$ limit of equations \eqref{saddle} for $x\notin(-\infty,\gamma]\cup\text{supp}\ \rho_+$ can then be written as

\begin{subequations}
	\label{eom}
	\begin{align}
		\label{eom1}
		U_+^{\prime}(x)&=x_3^*(x)+w_+(x)+w_+(-x)\;,\\
		\label{eom2}
		U_3^{\prime}(x)&=x_+^*(x)+w_3(x)\;.
	\end{align}
\end{subequations}

\noindent These equations determine the desired disc partition functions in terms of the multi-valued function $x_3(x)$ and its inverse: $w_3(x)$ follows straightforwardly from \eqref{eom2}, and $w_+(x)$ may be obtained as the solution to the Riemann-Hilbert problem defined by properties \ref{prop1} and \ref{prop2}, with the condition that $w_+(x)=x^{-1}+\mathcal{O}(x^{-2})$ as $x\to\infty$ as a consequence of the definition \eqref{w+}. 

\begin{figure}[b]
	\sidecaption
	\begin{pspicture}(7,4)
\psset{unit=.51cm}

\psline(0,0)(5,0)
\psline(0,1)(5,1)
\psline(0,2)(5,2)
\psline(0,3)(5,3)
\psline(0,4)(5,4)

\psline(1,0)(1,1)
\psline[doubleline=true](4,1)(4,2)
\psline[doubleline=true](1,2)(1,3)
\psline(4,3)(4,4)

\rput[tl](5.5,0.5){$x_3^{(1)}\propto-\sqrt x$}
\rput[tl](5.5,1.5){$x_3^{(2)}\propto \sqrt x$}
\rput[tl](5.5,2.5){$x_3^{(3)}\equiv x_3^*\propto x^2$}
\rput[tl](5.5,3.5){$x_3^{(4)}\propto x$}
\rput[tl](5.5,4.5){$x_3^{(5)}\propto x$}

\end{pspicture}
	\caption{Analytic structure of the function $x_3(x)$. Horizontal lines depict sheets, vertical lines cuts. Of the latter, double lines correspond to finite cuts on the real axis and single lines to cuts that extend to $\infty$}	
	\label{fig:sheets}
\end{figure}
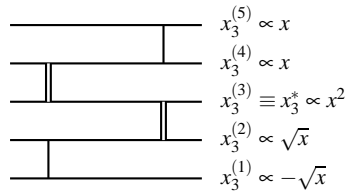

It remains to determine the function $x_3(x)$. To this purpose, note first that property \ref{prop3} allows us to write \eqref{eom2} as $U_3^{\prime}(x_3(x))=w_3(x_3(x))+x$, which when expanded about $x=\infty$ gives the asymptotic behaviour $\pm\sqrt{x}$ of $x_3(x)$ on the initial sheet \eqref{fcts} and another sheet connected to it through $(-\infty,\gamma]$. We may then use \eqref{eom1} to determine the cut structure and asymptotic behaviour on all other sheets by circling around the various branch points. The result of this procedure is depicted in Fig. \ref{fig:sheets}. We thus make an ansatz that the function $x_3(x)$ takes values on an algebraic curve\footnote{Closely related generating functions were proven to satisfy algebraic equations in \cite{Bernardi11}. We expect similar theorems to hold in our case.} 
$\mathcal{C}$,

\begin{eqnarray}
	\label{curve}
	\mathcal{C}=\{(x_3,x_+)|Q(x_3,x_+)=0\}\;,\\
Q(x_3,x_+)=\prod_{k=1}^{5}(x_3-x_3^{(k)}(x_+))\;.
\end{eqnarray}

\noindent Demanding that the coefficients of $x_+$ themselves be polynomials in $x_3$ fixes $Q(x_3,x_+)$ up to 10 unknown functions of the coupling constants. These unknowns are fully determined if we restrict to solutions for which the genus of the curve vanishes
: in that case, we may parametrise $\mathcal{C}$ by two rational functions $\mathbb{CP}^1\setminus\{z_i\}\ \to\ \mathcal{C}$. The poles $z_i$ of these functions will correspond to the asymptotic regions on each sheet of $x_3(x)$. For concreteness, we may position these poles at the canonical points $0$, $1$ and $\infty$ using a conformal transformation of $z$. A possible parametrisation then reads

\begin{equation}
	\label{param}
	x_+(z)=\frac{\sum_{k=0}^5\alpha_k z^k}{z^2(z-1)}\;,\qquad x_3(z)=\frac{\sum_{k=0}^5\beta_k z^k}{z(z-1)^2}\;.\\
\end{equation}

\noindent These functions are single-valued on the punctured Riemann sphere and cover $\mathcal{C}$ exactly once. Demanding that their Laurent expansion about each pole reproduce the appropriate asymptotic behaviour, we obtain 4 conditions per pole allowing us to solve for the 12 unknowns $\alpha_k$, $\beta_k$, $k=0\dots 5$. This completes our solution for the disc partition functions with fixed and mixed boundary conditions.

\section{Critical Behaviour}
\label{critical}

\noindent To explore the phase diagram of the model, we do not need to solve for the coefficients in \eqref{param} explicitly. Instead, we note that the possible critical exponents are determined by the multiplicity of the singularity at the left edge of the spectral density, which controls the large-order behaviour of the generating function $w_+(x)$. As we let $g$ and $c$ approach their critical values, the various branch points on the curve merge so that at the highest critical point $\mathcal{C}$ will exhibit a single singularity. Its multiplicity is fixed to 5 if the genus of $\mathcal{C}$ vanishes. This results in the conditions

\begin{equation}
	\frac{\partial^{m+n}}{\partial x_+^m\partial x_3^n}Q(x_3,x_+)=0\;, \qquad m+n<5\;,
\end{equation}

\noindent giving the critical values $c_c=2+\sqrt{47}$, $g_c=\sqrt{105}/2$ in agreement with \cite{ZJ99,Daul94}; the singularity is located at $x_{+,c}=0$, $x_{3,c}=-g_c^{-1}(4+c_c)/2$. To find the scaling behaviour of $w_+(x)$ near this point, it is useful to resolve the branch points at $\gamma$ via the change of variables $x(\zeta)=\gamma(1-2\zeta^2)$. We also introduce the auxiliary function

\begin{equation}
	f(\zeta)=P(\zeta)-\int_{\zeta(\text{supp}\ \rho_+)}\D \zeta^{\prime}\frac{\rho_+(x(\zeta^{\prime}))}{\zeta-\zeta^{\prime}}\;,\qquad \zeta\notin\zeta(\text{supp}\ \rho_+)\;,
\end{equation}

\noindent where $P(\zeta)$ is a polynomial of degree 4 and we integrate over the image of the support of $\rho_+$ obtained by selecting the positive branch of $\zeta(x)$. For a suitable choice of coefficients in $P$, \eqref{eom1} then has the equivalent homogeneous form

\begin{equation}
	\label{eom3}
	2\operatorname{Re} f(\zeta)+f(-\zeta)+f(\sqrt{1-\zeta^2})+f(-\sqrt{1-\zeta^2})=0\;,\quad \zeta\in\zeta(\text{supp}\ \rho_+)\;.
\end{equation}

\noindent If we parametrise the vicinity of the singularity as $\zeta\propto\cosh\phi$ so that $f(\zeta)\propto\cosh(\mu\phi)$, we find that \eqref{eom3} implies $5\mu=\pm 4 n+20 m$ with $n\in\{1,2\}$ and $m\in\mathbb{Z}$. Taking into account the upper bound on $\mu$ implied by the degree of $\mathcal{C}$ leads to the conclusion that $\mu=12/5$ for $g=g_c$, $c=c_c$ and hence that at the critical point, $w_+(x)$ has scaling exponent $6/5$. Indeed, the corresponding value of the string susceptibility $\gamma_s=-1/5$ agrees with that found for fixed boundary conditions \cite{ZJ99,Daul94,Eynard99}.

\section{Discussion}
In this contribution, we computed the partition function for the 3-states Potts model on the randomly triangulated disc with both fixed and mixed boundary conditions. By requiring that the large $N$ saddle point equations possess a consistent analytic continuation, we found that both boundary conditions can be encoded in a single algebraic curve and determined the scaling behaviour of the disc partition functions at the highest critical point from its singularities. As expected, the resulting value of the string susceptibility $\gamma_s=-1/5$ is independent of the chosen boundary conditions and consistent with Liouville quantum gravity coupled to conformal matter with central charge $c=4/5$. Given the symmetries of the model, the identification of the latter sector with the $(A_4,D_4)$ modular invariant hence appears justified.

The algebraic curve \eqref{curve} forms part of the initial data for topological recursion \cite{Eynard09} which enables systematic computation of finite $N$ corrections to various observables. Of particular interest would be the computation of cylinder amplitudes between different boundary states to probe the spectrum of the continuum theory beyond the planar limit. This would allow a check of the conjecture made in \cite{Kawamoto08} that in the presence of gravity, not all states in Table \ref{tab:states} are independent\footnote{Similar observations were made for the $(A_p,A_q)$ minimal models coupled to gravity \cite{Seiberg04,Atkin10,Atkin12}.}. 
We intend to address these and related issues in a forthcoming publication.
\label{discussion}

\begin{acknowledgement}
	BN would like to acknowledge support by the German National Academic Foundation and STFC grant ST/J500641/1.
\end{acknowledgement}

\bibliographystyle{spphys}
\bibliography{main}
\end{document}